\documentstyle[epsf]{mn}

\def\rx{1WGA J0053.8--7226\ }
\def\rxj{1WGA J0053.8--7226}
\def\ecms{ergs~cm$\rm ^{-2}~s^{-1}$}
\def\kms{km~s$^{-1}$\ }
\def\km{km~s$^{-1}$}
\def\lap{$\buildrel < \over {_\sim}$} 
\def\ap{${\sim}$}
\def\er{${\pm}$}
\def\gradi{\ifmmode{^\circ}\else$^\circ$\fi}

\def\degmark{^\circ}

\begin{document}

\title[The hard transient \rx]{Multiwaveband studies of the hard {\it
ROSAT} SMC transient \rxj: a new X-ray pulsar}

\author[D. A. H. Buckley et al.]
{D. A. H. Buckley$^{1}$, M. J. Coe$^{2}$, J. B. Stevens$^{2}$, K. van
der Heyden$^{1}$\cr
L. Angelini$^{3}$, N. White$^3$ and P. Giommi$^3$\\
$^{1}$South African Astronomical Observatory, P.O. Box 9, Observatory 7935,
Cape Town, South Africa\\
$^{2}$Physics and Astronomy Dept., The University, Southampton, SO17 3BJ, UK \\
$^{3}$ HEASARC, NASA/Goddard Space Flight Center, Greenbelt, MD 20771, USA\\
}

\date{Accepted \\
Received : Version November 1997 \\
In original form ..}

\maketitle

\begin{abstract} 
We report on two optical candidates for the counterpart 
to an X-ray source in the SMC, \rxj,
identified as a serendipitous X-ray source from the {\sl ROSAT}
PSPC archive, and also observed by the {\sl Einstein} IPC.
Its X-ray properties, namely the hard X-ray spectrum,
flux variability and column density indicate a hard, transient source, with
a luminosity of \ap 3.8 $\rm \times 10^{35} ergs~ sec^{-1}$. 
{\it XTE} and {\it ASCA} observations have confirmed the source to be
an X-ray pulsar, with a 46 s spin period.
Our optical observations
reveal two possible candidates within the error circle. Both exhibit
strong H$\alpha$ and weaker H$\beta$ emission. The optical colours
indicate that both objects are Be-type stars.
The Be nature of the stars implies that the 
counterpart is most likely a Be/X-ray binary system. Subsequent IR photometry 
({\it JHK}) of one of the objects 
shows the source varies by at least 0.5 mag, while the {\it
(J-K)} measured nearly simultaneously with the {\it UBVRI} and spectroscopic 
observations indicate an IR-excess of \ap 0.3 magnitudes.

\end{abstract}


\section{Introduction}
Accretion onto a compact object invariably results in X-ray emission, with
the luminosity determined by such important parameters as the size and mass of the
accreting object, and the mass accretion rate. Binary stars represent the
best laboratories for the study of such accretion processes, where the
accreting star can be a white dwarf, neutron star or a black hole. Whereas
known white dwarfs in cataclysmic variables are generally low luminosity X-ray
sources, and relatively nearby (most \lap 1 kpc), the neutron star sources
are often extremely luminous, and detectable at extragalactic distances.

Most (\ap 80\%) of the higher mass systems (High Mass X-ray Binaries, or 
HMXBs), which consist of early spectral type (O or B) stars losing mass to 
their neutron star companions, are Be/X-ray or Supergiant binaries. The
orbits of these binary systems are generally wide and eccentric, with
periods in the range 0.7/1.4d for RX J0050.7-7316 (Coe \& Orosz, 2000) to
262d for SAX J2239.3+6116 (in't Zand et al, 2000). Usually the only optical
signature of these stars is Balmer emission, and in that respect alone it is
often difficult to ascertain the X-ray nature of these systems. With Be stars
being relatively common in the Galactic plane, it has only been possible to
confirm Be counterparts to X-ray sources using imaging X-ray instruments,
as provided by such satellites as {\it Einstein} (HEAO-2) and {\it ROSAT}. 

Much progress in understanding the physics of these systems has resulted
from multiwaveband campaigns, spanning the
IR--optical--UV--X-ray--$\gamma$-ray domains. Observations of the Be star in
the optical and IR sheds light on the physical conditions under which the
neutron star accretes. In addition, with contemporaneous X-ray observations,
which provide direct information on the accretion process, it is possible
to investigate the correlation between the mass loss rate from
the Be star and the accretion rate. Thus it is possible, through long-term
multi-wavelength programs, to build a more complete picture of the whole 
accretion process, and to study it as a function of time.

In this paper we report on two possible counterparts to a hard X-ray
transient source in the SMC, \rxj, which is most likely a new member
of the Be/X-ray binary subclass of HMXBs. The {\it ROSAT} position
coincides with a previously know hard spectrum source from HEAO-2
({\it Einstein Observatory}) IPC pointings.  Our optical/IR
observations of one of the candidates reveal IR variability of \ap 0.5
mag and persistent Balmer emission, with a redshift consistent with
SMC membership.  The UBVRI colours and optical spectra of both
candidates indicate an early B-spectral type.

In 1997 November, following our observations, an X-ray source identified as
XTE J0053-724 went into outburst and revealed a 46 s X-ray pulsation (Corbet
et al. 1998). Though originally, and incorrectly, identified as SMC X-3,
subsequent ASCA observations in 1997 December revealed that the pulsating
source was actually 8' away from SMC X-3, establishing it to be a new X-ray
pulsar. Its new ASCA-determined position is coincident with the previously 
catalogued X-ray source \rx and the data we present here confirms this 
association and provides convincing evidence that it is a Be/neutron star 
binary.

\section{ROSAT observations}

The WGA catalogue (White, Giommi, Angelini 1994)
was generated using all pointed ROSAT PSPC data available in the archive up to
1994 November and its first revision includes data up to 1995 March. The
catalogue was created using the rev1 ROSAT data files, where data taken
several months apart belonging to the same sequence were combined in one
file. A variability test for each detected source was included as part of
the processing. The method used compares the time arrival distribution of the
photons collected in each pixel with the corresponding distribution of the
entire image using a Kolmogorov-Smirnov (KS) test (Giommi, Angelini, White
1995). The transient nature of \rx was discovered using this
method, in the sequence 600195, which included data from 1991 October and
1992 April. The source was detected in 1992 April at coordinates R.A.=00$^{\rm
h}\, 53^{\rm m}\, 53.8^{\rm s}$ and Dec=$-72\degmark\, 26' 35''$ (J2000).
Positional errors from the WGA positions vary from $13-50''$, depending how
far the source is off-axis.

After discovering the transient behaviour, the ROSAT public archive
was searched for additional PSPC and HRI public data containing the
source position.  Table 1 gives the observation log \footnote{ We used
the rev 2 data processed with SASS7, which included among other
changes, a better attitude reconstruction, and the splitting of the
sequence into separate files when the guide stars change (see also:
Pisarski 1995).}  together with the nominal exposure, the off-axis
position of the source in the HRI or PSPC Field of View measured from
the nominal pointing, and the count rate.

A count rate (or upper limit) is given in Table 1 for most of the
observations with the following caveats. For a large off-axis angle,
the point spread function (psf) is very large: at 50 arcmins the 50\%
power radius is $\sim 3$ arcmins (see ROSAT Handbook). That far
off-axis, the source position is typically at the detector boundary
and the optimized radius falls outside the detector image, resulting
in an underestimate of the countrate. \rx is at $\sim $~5 arcmin
from another transient RX J0054.9-7210 (Kahabka and Pietsch
1996), which is also an {\it Einstein} IPC source: No. 9 in Bruhweiler
et al (1987) and No. 35 in Wang and Wu (1992). At such large off-axis
angles, if this second source was active, the optimized radius to
evaluate the count rate for \rx is contaminated by emission from
RXJ0054.9-7210.  For these observations no count rate or upper limits
are given in Table 1, although a column indicates whether \rx was
detected (`on' or `off').

\begin{figure}
\begin{center}
{\epsfxsize 0.99\hsize
\epsffile{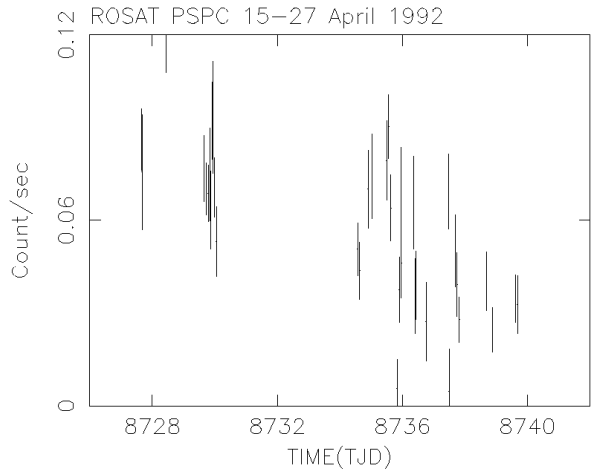}}
\end{center}
\caption{The background subtracted lightcurve is obtained for the PSPC
observations 600195a01 and 600196a01 for the 1992 April observation. The data
are corrected for PSF and vignetting effects. Each bin corresponds to 1000
seconds.}
\end{figure}

The source is highly variable and it was detected in 1992 April (PSPC),
1993 October (PSPC), and 1996 June (HRI). Although the coverage is not
complete, the outbursts for the 1992 April and 1993 October observations last
more than a month. Figure 1 shows the background subtracted lightcurve for
the 1992 April outburst combining data from the sequence 600195 and 600196,
where only the declining phase is seen. The PSPC spectrum obtained for the
1992 April observation (only when the source is at $\sim 20$ arcmin
off-axis, because of the response uncertainties at large off-axis angles) is
shown in Figure 2. We fit the spectrum using a power law and absorption. The
best fit parameter values are a spectral index of 1.6 with an equivalent
hydrogen absorption of $3 \times 10^{21}$~cm$^{-2}$ (shown in Figure 2), but
these two parameters can not be simultaneously constrained. 

A model with a fixed absorption at the galactic value in the SMC
direction ( $\sim 6 \times10^{20}$~cm$^{-2}$) and a power law,
overestimates the low energy part of the spectrum. If 1WGA
J0053.8-7226 is a Be star, as argued later, it is reasonable to assume
that the outburst spectrum should be similar to that of 1WGA
J0051.8-7231, another Be transient in the SMC (Israel et al. 1997). Fixing
the spectral index to 1.1, as for 1WGA J0051.8-7231, we obtained an
N$_H$=$(2.3 \pm 1.1) \times 10^{21}$~cm$^{-2}$. This value seems
consistent with the contour map in Kahabka \& Pietsch (1996) where \rx
is close to source number 103 in their paper.  This gives a 0.1-2.0
keV unabsorbed X-ray flux of $7.5 \times 10^{-13}$~erg~cm$^{-2}~\rm
s^{-1}$ which corresponds to a luminosity of $\sim 3.8 \times 10^{35}$
erg~s$^{-1}$ for a distance of 65 kpc.

\begin{figure}
\begin{center}
{\epsfxsize 0.99\hsize
\epsffile{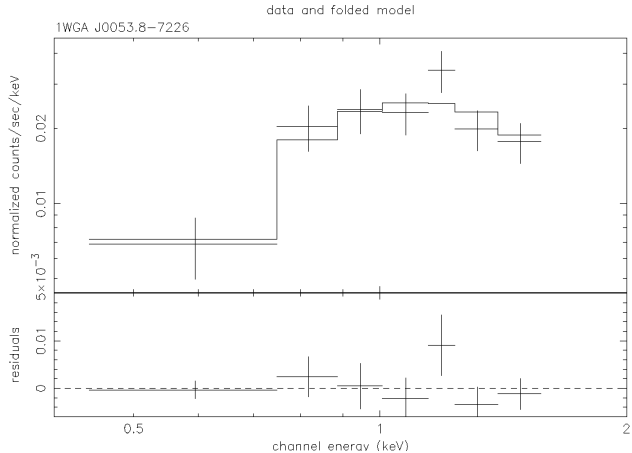}
}\end{center}
\caption{X-ray spectrum, power law fit and residuals for the 600195a01 
observation of\rxj.}
\end{figure}

The 1996 June HRI observation allowed us to derive a better position
because of the higher HRI angular resolution compared to the PSPC, and
because the source location is at a smaller off-axis angle ($\sim 3$
~arcmins) compare to the PSPC observations. At equinox 2000 this is
R.A.=00$^{\rm h}\, 53^{\rm m}\, 55.6^{\rm s}$, Dec=$-72 \degmark\, 26'
44''$ with a conservative error of $15''$ (see also HRI report, Harris
(1996)).

The position obtained from the PSPC No. 3 observation using the rev2
data is R.A.= 00$^{\rm h}\, 53^{\rm m}\, 56.0^{\rm s}$ Dec=$-72 \degmark\,
26'\, 34.8''$ (\er $25''$). The 1WGA J0053.9-7226 position is consistent with 
the {\it Einstein} IPC source No. 8 in Table 2 of Bruhweiler et al (1987) and
source No. 34 in Wang and Wu (1992). The coordinates of the {\it
Einstein} source from Wang and Wu (1992) are RA=$00^{\rm h}\, 53^{\rm m}\,
54.2^{\rm s}$ Dec=$-72\degmark\, 26'\, 27''$ in equinox 2000, with an error
\er $40''$.

\begin{table*}
\begin{center}
\begin{minipage}{160mm}
\caption{Table~1: Observation log of 1WGA J0053.8-7226}
\begin{tabular}{lrllrcrll}
\hline \hline \\
\multicolumn{1}{l}{No.} & \multicolumn{1}{c}{Instrument} &
\multicolumn{1}{c}{Year} & \multicolumn{1}{c}{Day/Month} &
\multicolumn{1}{r}{Exp.$^{1}$} & 
\multicolumn{1}{c}{Seq.} & \multicolumn{1}{c}{Off-axis} & 
\multicolumn{1}{c}{On/Off} & \multicolumn{1}{l}{Count rate $^{2}$}\\   
   &            &      &  & (s) &      & (arcmin)   &        
& ($\times 10^{-2}$)     \\
\hline \\[-3mm]
1  &PSPC    & 1991 & 8/Oct 3/Nov & 16644 & 600195A00 & 21.9 & off  &  $ < 0.536$ \\
2  &PSPC    & & 9/Oct 2/Nov     & 1303  & 600196A00 & 49.3 & off$^b$ & -- \\
3  &PSPC    & 1992 & 15-25/Apr  & 22223 & 600196A01 & 49.3 & on   & $6.14 \pm0.38$ \\
4  &PSPC    & & 17-27/Apr      & 9443  & 600195A01 & 21.9 & on   & $5.36 \pm0.38$  \\
5  &HRI     & & 23-24/Oct      & 1096  & 400237N00 & 10.3 & off  & $ <0.844$   \\
6  &PSPC    & & 6/Dec          & 3561  & 600455N00 & 50.5 & off$^a$ &  -- \\
7  &PSPC    & 1993 & 29-30/Mar & 5214  & 400300N00 & 55.0 & off$^{a,c}$& -- \\
8  &PSPC    & &10-25/Apr     & 14207 & 600452N00 & 54.7 & off$^{a,c}$& -- \\
9  &PSPC    & & 16-22/Apr      & 1721  & 600455A01 & 50.5 & off$^{a,c}$& -- \\ 
10 &HRI     & & 17/Apr/93      & 1167  & 400237A01 & 10.3 & off  & $ < 0.769$ \\
11 &PSPC    & & 10/May/93      & 17593 & 600453N00 & 19.2 & off  & $ < 0.828$ \\
12 &PSPC    & &712-13/May      & 4902  & 500142N00 & 52.6 & off  & $ < 3.91$ \\
13 &PSPC    & & 1-14/Oct       & 16663 & 600452A01 & 54.7 & on   & $ 9.50 \pm0.70$ \\
14 &PSPC    & & 1-9/Oct        &  7199 & 400300A01 & 55.0 & on   & $ 8.4 \pm1.5$ \\
15 &PSPC    & & 7-10/Oct       & 4595  & 600455A02 & 50.5 & on   & $ 16.4 \pm1.6$ \\
16 &PSPC    & & 14-29/Oct      & 20845 & 500250N00 & 53.5 & on   & $ 9.95 \pm0.55$ \\
17 &HRI     & 1994 & 1/Apr     &  1301 & 400237A02 & 10.3 & off  & $ < 0.701$ \\
18 &PSPC    & & 4-5/May        & 4129  & 600455A03 & 50.5 & off  & $ < 0.395$ \\
19 &PSPC    & & 5/May          & 4111  & 400300A02 & 55.0 & off  & $ < 3.96$  \\
20 &HRI     & 1996 & 26/Apr    & 2006  & 600810N00 & 3.2  & off  & $ < 0.451$ \\
21 &HRI     & & 10/June        & 4720  & 600810N00 & 3.2  & on   & $ 4.06 \pm0.32$ \\ 

\hline 
\end{tabular}

\noindent{Table Notes:} \\
$^{a}$ The nearby source (RX J0054.9-7210) is active \\
$^{b}$ The nearby source (RX J0054.9-7210) is active. The source state is 
derived from the nearby observation at smaller off-axis angle. \\
$^{c}$ The source location is at the detector boundary, but no excess is seen \\
$^{1}$ The exposures listed are the nominal values. Close to the PSPC support 
structure the exposure is much reduced. The count rates are derived using an 
exposure map.\\
$^{2}$ Upper limits are given at 3 sigma\\ 

\end{minipage}
\end{center}
\end{table*}

The source was noted by Bruhweiler et al. to have the hardest spectrum 
amongst the 35 listed, and assumed to be a stellar source.  From an IPC 
count rate of 0.0088 $\rm cts~ s^{-1}$, Bruhweiler et al. determined an
X-ray luminosity of 1.8 $\rm \times 10^{35} ergs~ s^{-1}$ and hardness ratio 
$HR = (H - S)/(H + S)$ = (0.8 to 3.5 keV -- 0.2 to 0.8 keV) / 
(0.8 to 3.5 keV + 0.2 to 0.8 keV) = 0.76, consistent with the hard spectrum we
report here.

\section{Optical identification}
 
The field of \rx is shown in Figure 3. There are 4 stars lying within
the X-ray error circle labelled as Stars A, B, C and F. Star A is, in
fact, a double and both components were subsequently observed on
1995 September 22/23, using the SAAO 1.9-m telescope with the
Cassegrain spectrograph. Spectra in the interval 3300-7500\AA, at a
resolution of \ap 7\AA, were obtained using the Reticon Photon
Counting System (RPCS). Because the two components of Star A were
separated by only a few arcsec in the E-W direction, it was impossible
to get separate spectra of the two. The combined spectrum was clearly
very blue, and with a strong H$\alpha$ emission line (E.W. =
--13\AA). Although the RPCS is only a 1D detector, it was possible to
determine which of the two stars was responsible for the Balmer
emission. This was done by driving the stellar image off to one
extreme of the slit decker, thereby excluding most of the light of the
companion. The exercise was then repeated for the other star. It was
therefore possible to obtain two spectra, whose major light
contribution was one or other of the two stars.  From these
observations we were able to show that the fainter, easternmost, star
was the H$\alpha$ emitter. This has been subsequently confirmed from
both optical photometry and CCD spectroscopy, reported later in this
paper.

\begin{figure}
\begin{center}
{\epsfxsize 0.99\hsize
\epsffile{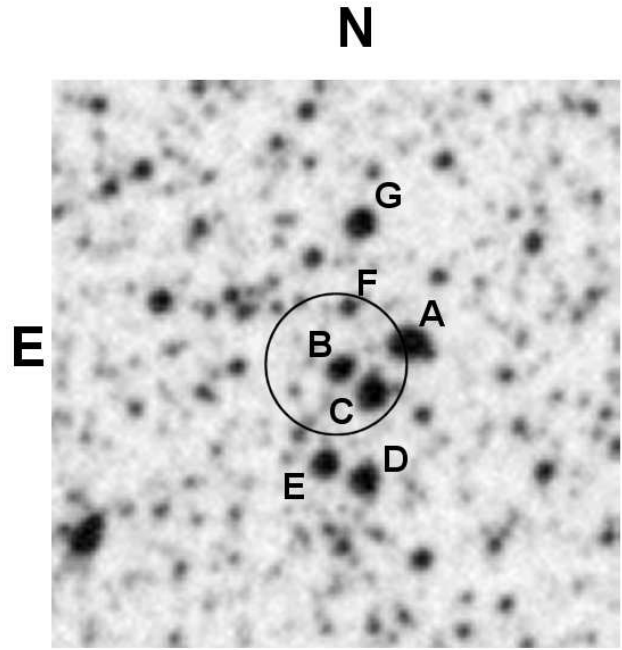}
}\end{center}
\caption{Finding chart (\ap 2 $\times$ 2 arcmin) for the \rxj ~field 
from the Digital Sky Survey (http://archive.stsci.edu/dss)
showing all the possible counterparts
referred to in the text. The HRI 90\% confidence error circle (radius
= $15''$) is marked, the centre of which is $\rm R.A. = 00^H 53^M
55.6^S Dec = -72^{\circ} 26' 44''$ (Equinox 2000).  }

\end{figure}

\begin{table*}
\noindent
\begin{minipage}{180mm}
\caption{Optical magnitudes/colours of the two main optical candidates for \rx on
1995 Nov 25/26}
\begin{tabular}{|lcccccrrrrr|}

Star	& U & B	& V & R	& I & (U--B) & (B--V) & (V--R) & (V--I) & (R-I) \\
\hline

Be star component of A & 14.004 & 14.876 & 14.927 & 15.000 & 14.995 & --0.87 & --0.05 & 
--0.07  & --0.07 & 0.01 \\
	& (\er 0.011)	& (\er 0.016)	& (\er 0.023)	& (\er 0.024)	&
(\er 0.024) & (\er 0.03) & (\er 0.04) & (\er 0.05) & (\er 0.05) & (\er 0.05)\\
F-star component of A &  14.553 & 14.444 & 14.075 & 13.730 & 13.412 & 0.109 & 0.369 &   
0.346 & 0.664 & 0.318 \\
& (\er 0.016) & (\er 0.007) & (\er 0.006) & (\er 0.005) & (\er
0.005) & (\er 0.023) & (\er 0.013) & (\er 0.011) & (\er 0.011) & (\er 0.010)
\\
Star B &  13.833 & 14.637 & 14.881 & 14.725 & 14.711 & -0.80 & -0.25 &   
-0.16 & 0.17 & 0.01 \\
& (\er 0.016) & (\er 0.007) & (\er 0.006) & (\er 0.005) & (\er
0.005) & (\er 0.023) & (\er 0.013) & (\er 0.011) & (\er 0.011) & (\er 0.010)
\\
\hline

\end{tabular}
\end{minipage}
\end{table*}

CCD photometry of the field of \rx was made using the SAAO 1.0-m telescope 
on 1995 November 26/27. The observations were made using the Tek8 CCD camera,
employing $UBVR_CI_C$ and H$\alpha$ filters, at a plate scale of 0.35 
arcsec $\rm pixel^{-1}$. All observations were obtained in photometric 
conditions, and in good seeing (FWHM \ap 1"), with exposure times of 600 s 
({\it U}), 300 s ({\it B}), 200 s ($V, R_C~ \&~ I_C$) and 1000 s (H$\alpha$).

All of the CCD frames were reduced in a similar manner, with bias removal,
flat-fielding, and subsequent reductions using {\it DoPHOT} (Mateo and
Schechter 1989). Observations of E-region standard stars (Menzies et al.
1989) were used to determine aperture magnitude zero points. Both the PSF
fit and aperture magnitudes were determined for all stars in the frames, and
transformed to the natural $UBVR_CI_C$ system. Colour-magnitude and colour-colour
diagrams were made in order to search for stars of unusual colour. In the
past the H$\alpha$ filter has been used successfully in conjunction with the
{\it R}-band filter, to search for H$\alpha$ emission line objects. No stars
on the frame appeared to show significant deviations suggestive of
very strong
H$\alpha$ emission.

Our CCD observations resolve the two components of Star A
and indicate that they are separated by 2.1 arcseconds in
an E-W direction. We give the photometric results in Table 2 for the
stars labelled A and B in Figure 3.

We find that Star B and one component of Star A are very blue, and lie
with a group of several other blue stars off the normal locus for
stars in the two-colour $(U-B) vs (B-V)$ diagram, implying some
reddening. An $E(B-V)$ of 0.15--0.20 would move the stars back onto
the normal locus for early B-stars in the two colour diagram, namely a
$(B-V)_0$ \ap --0.20 to --0.25. The fainter component of Star A has a
$(B-V)$ colour of 0.37 and is consistent with an early F spectral
type, while the $(V-R)$~ and $(V-I)$~ colours imply a somewhat later
type (F8). Our CCD spectra (Section 5) are more consistent with the
earlier spectral type, so we adopt F2 for the this star's type.

\section{Followup optical and IR photometry}

Subsequent optical (${\it BVR_C}$) photometry was performed on 1996 October 
5/6. These CCD observations were also undertaken on the SAAO 1.0-m
telescope, using the same (Tek8) camera, but employing the 3$\times$
``Shara'' focal reducer (kindly provided by Dr M. Shara, STScI), giving an
effective field size of 9 $\times$ 9 arcmin, with a scale of 1.05 arcsec
$\rm pixel^{-1}$ (this was utilised for a parallel program of optical
identifications necessitating a larger field coverage). No {\it U}-band 
observations were possible, due to the poor throughput of the focal reducer 
at this wavelength. Because of the plate scale, the close pair of
objects labelled Star A in Figure 3
were not well resolved. In fact only for the {\it B} and {\it V}
frames was it possible to derive PSF fits to both stellar images, although
the results are far from reliable.
The aperture magnitude of both stars (i.e. using an aperture large enough to
include both) was V = 13.64, compared to a combined V = 13.667 for 1995 
November photometry, indicating very similar combined flux,
and therefore little optical variability. The individually fitted PSF 
magnitudes were rather different, however, no doubt a consequence of severe
undersampling of the images, plus the fact that there are in fact 2
fainter stars within several arcseconds (Star C and F), 
i.e. {\it four} stars in total.
Indeed our 1995 November observations resolved all 4 stars.

IR $(JHK)~$ photometry of Star A was performed on 1995 November 14/15 and 1996 
October 2/3 using the MkIII IR photometer on the SAAO 1.9-m telescope.
A chopping secondary defines two effective apertures, one containing the
source, the other background. After a short 10 sec integration, the
telescope ``nods'' such that the source is within the aperture previously
defined as background. Thus two ``off-source'' positions are used, in order to
minimise any systematic errors. Observations are repeated, alternately
nodding to the alternative background positions, until a sufficient S/N
ratio is obtained. Most of the observations consisted of 5 observation
modules of 40 s integrations, or a total of 200 s.
For both sets of observations, aperture sizes of 6 and 9 
arcsecond diameter were used, respectively, for the 1995 and 1996 observations. 

The {\it JHK} magnitudes are therefore of the {\it combined} light of both 
objects representing 
Star A (in fact also including the two other fainter, by 2--3 mags at V,
neighbouring stars), and they are given in Table 3 with their standard
internal errors. We were careful to observe a nearby JHK standard close in
time (\ap 40 min) to the \rxj ~observation. Therefore the true errors should
not be too different to those quoted. 
Included are the 
deconvolved magnitudes for both components of Star A. 
These deconvolutions were made assuming 
that the IR colours, {\it(V--K), (H--K)} and {\it(J--K)}, of the F-star could 
be determined on the basis of the observed {\it(B--V)}, using the 
relationships between the SAAO IR {\it JHK} and Johnson-Cousin's $UBVR_CI_C$ 
systems, determined by Caldwell et al. (1993). For the colours in question, 
these relationships are independent of luminosity class.
From the observed {\it V~} and predicted IR colours ({\it V--K, 
H--K, J--K}) for the F-type contaminating star, we derived the {\it JHK~}
magnitudes of both components of Star A.

\begin{table}
\noindent
\begin{minipage}{85mm}
\caption{IR magnitudes for Star A and its nearby companion}
\begin{tabular}{|lcccc|}

Star		& Date 		& J 		& H		& K\\
\hline
{\it Combined:}	& Nov 95	& 13.18		& 12.93 	& 12.84\\
		&		& (s.e. 0.05)	& (s.e. 0.05)	& (s.e. 0.07) \\
		& Oct 96	& 12.69		& 12.47		& 12.45\\
		&		& (s.e. 0.10)	& (s.e. 0.07)	& (s.e. 0.05) \\
		& & & & \\
{\it Deconvolved:} & & &\\
Nearby F-star	& 		& 13.36		& 13.11		& 13.08\\
& & & & \\
Be star 	& Nov 95	& 15.22		& 14.97		& 14.60\\ 
Be star 	& Oct 96	& 13.53		& 13.35		& 13.34\\
\hline

\end{tabular}
\end{minipage}
\end{table}

For the 1996 observations, the derived {\it J--K} for the Be star
component of Star A is 0.19, and
assuming {\it E(J--K)} \ap 0.52 {\it E(B--V)}, leads to a value of $({\it
J-K})_0$ = 0.11 (for {\it E(B--V)} = 0.15). This value can be compared to
that expected for a late-O/early-B star (\ap --0.2), indicating that \rx
exhibits an IR excess of \ap 0.3 mags. Such an excess is a common feature of
Be/X-ray binaries (e.g. Coe et al. 1994), and is probably related to the
amount of mass present in the Be star's circumstellar disc.

\section{Optical spectra}

Our discovery spectrum, taken with the RPCS, lead to the
identification of the Balmer emission (in H$\alpha$ and weakly in
H$\beta$) in the eastern component of the close pair labelled Star A
in Figure 3.  However, the RPCS was unable to obtain spectra of both
components separately.  This was subsequently achieved using the {\it
SITe2} CCD detector on the Cassegrain spectrograph, with the same
grating and wavelength coverage. The observations (2 $\times$ 600 s)
were kindly undertaken by H. Winkler on 1997 October 2/3, in
photometric conditions. The flux standard star EG 21 was observed
immediately after the \rx observations, and the data reduced using
{\bf IRAF}\footnote{IRAF is distributed by NOAO, which is operated by
AURA, Inc., under contract to the NSF.}  tasks in {\bf ccdproc,
onedspec} and {\bf twodspec}.

The close proximity of the two objects in Star A meant that the
individual spectra were hardly resolved in the spatial direction
(i.e. perpendicular to the slit) on the CCD, having a typical cross
section of FWHM \ap 3 pixels. Using the {\bf apall} program, we
experimented extracting the spectrum using a range of apertures. The
H$\alpha$ emission was clearly visible over several CCD rows, which we
used as an initial indicator for which rows to extract for the Be star
spectrum. The best results were achieved by optimally extracting the
spectra over a range of rows, starting 6 pixels to one side of the
peak in the cross section, to within 1 pixel of this peak (or 5 pixels
in total). Likewise, the contaminating nearby F-star was extracted
over the remaining pixels in the profile, i.e. 6 pixels on the other
side of the peak to the limit of the Be star extraction (7
pixels). The spectra were first summed over 30 pixels at a time in the
dispersion direction, to obtain the centroid (in the spatial
direction) of the spectrum, and then traced using a cubic spline.

Observations of Star B were subsequently carried out using the same
telescope and instrument set up on 19/20 August 1998.

Our results are shown in Fig. 4 where the spectra of both Stars A and
B are shown. The spectrum of the F star near to Star A is shown in
Figure 5.  Although sky background has been subtracted, there has been
no attempt to remove the telluric lines (e.g. the A-band). The nearby
F-star shows strong Ca H \& K and Balmer absorption lines, on a blue
continuum. Judging from the relative strength of the G-band with
respect to H$\gamma$, an early F spectral type is implied, consistent
with the observed $(B-V)~$ of 0.37. The H$\alpha$ absorption line
maybe somewhat filled in from contamination by the Be component to
Star A. Indeed both spectra are likely to suffer from mutual
contamination at some level, although the results are far superior to
the RPCS spectra. We fitted Gaussian profiles to the emission lines
for Stars A and B in
both the RPCS discovery spectrum and the CCD spectra, and the results
are summarised in Table 4. The radial velocity values are consistent
with membership of the SMC, which has a mean systemic value of 166
$\pm$ 3 \km (Feast 1961). The same is true for the nearby F2 star,
which has a similar velocity shifted Balmer absorption lines.

\begin{table*}
\begin{center}
\noindent
\begin{minipage}{140mm}
\caption{Gaussian fits  to H$\alpha$ and H$\beta$}
\begin{tabular}{|llcccc|}
\hline
Object &Date 		& Radial Velocity 	& Flux		& E.W.	& FWHM \\ 
&		& (\kms)	& ( \ecms) & (\AA) & (\AA) \\
\hline
& & & & & \\
Star A&{H$\alpha$}: & & & \\
(Be star&22/23 Sep 95	& 177 		& 3.6 $\times 10^{-14}$	& --13.1 & 8.7 \\
component)&2/3 Oct 97	& 157		& 3.7 $\times 10^{-14}$	& --11.6 & 9.2 \\
&		& (\er 15)	& (\er 0.3 $\times 10^{-14}$) & (\er 0.9) &
(\er 0.7) \\
&	& & & & \\
&{H$\beta$}: & & & \\
&2/3 Oct 97	& 186  & 4.1 $\times 10^{-15}$ & --0.8 & 5.4 \\
&		& (\er 50)	& (\er 1.2 $\times 10^{-15}$ & (\er 0.2) &
(\er 2.1) \\ 
& & & & & \\
\hline
Star B&{H$\alpha$}: & & & \\
&19/20 Aug 98	&  -		& 5.8 $\times 10^{-14}$	& --15.3 & 10.7 \\
&		& 	& (\er 0.3 $\times 10^{-14}$) & (\er 0.9) &
(\er 0.7) \\
&{H$\beta$}: & & & \\
&19/20 Aug 98	& -  & 1.3 $\times 10^{-14}$ & --1.76 & 6.9 \\
&		& 	& (\er 1.2 $\times 10^{-15}$ & (\er 0.2) &
(\er 2.1) \\ 
\hline

\end{tabular}
\end{minipage}
\end{center}
\end{table*}

\begin{figure}
\begin{center}
{\epsfxsize 0.99\hsize
\epsffile{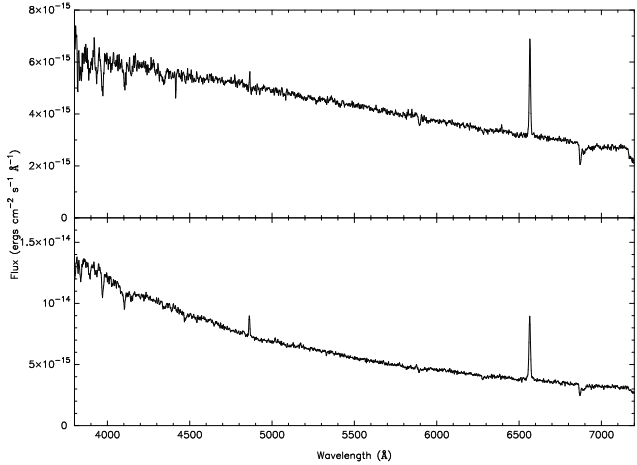}
}\end{center}
\caption{Top panel: spectrum of the eastern component of the 
star labelled as A in
Fig. 3. Lower panel: spectrum of Star B. }
\end{figure}

\begin{figure}
\begin{center}
{\epsfxsize 0.99\hsize
\epsffile{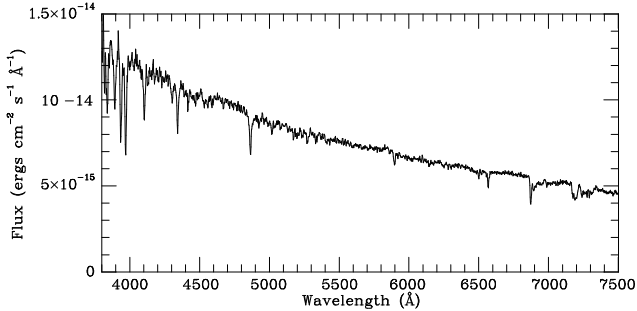}
}\end{center}
\caption{Spectrum of the F2 star which is the western component 
of the double labelled as A in Fig 3.}
\end{figure}

The remaining stars observed in, or near, the error circle are very
unlikely to be the counterpart to the X-ray source. Stars C and E are clear
A type stars, Star D is an F type, Star F a carbon star
and G is probably a late K type, probably dwarf star 
(and also a significant distance from the X-ray position).

\section{Discussion}

Apart from H$\alpha$, H$\beta$ and the other Balmer lines (seen in absorption
H$\gamma$, H$\delta$, H$\epsilon$, H8), there appears to be no other
significant features in the spectra of Stars A and B.  Regrettably the
S/N ratio of the spectrum is insufficient to obtain further spectral
typing information.  

However, for the Be component of Star A we can
use the contemporaneous optical and IR photometry from 1995 November
to model the spectral shape.  Assuming a value of E(B--V) = 0.15, we
de-reddened our data and fitted a Kurucz model atmosphere to the
photometry. Normalizing the model to our U band point results in an
acceptable fit for T = 22000 \er2000 K and log{\it g} = 4.0. This fit
is illustrated in Fig. 6, from which the IR excess over the stellar
model is clearly visible. This temperature range corresponds to a
B1-B2 spectral type.

Unfortunately we have no IR data for Star B to demonstrate that the IR excess
for that object also exists. However, the (U--B) and (B--V) 
optical colours are very similar to those of Star B indicating a
similar spectral type.

In the case of the Be star component to Star A membership of the SMC
is clearly established. Hence we can estimate the absolute magnitude
of this system using the observed apparent {\it V} magnitude in Table
2, correcting for an assumed absorption of {\it E(B--V)} \ap 0.2,
implying an $A_V~$ \ap 0.6. This gives $m_V$ \ap 14.3, and for a
distance modulus of $(m_V - M_v) = 18.9$ (Westerlund 1997), implies
$M_V$ = --4.6.  This suggests a luminosity class III-V. For the nearby
contaminating F2 star, the absolute magnitude is estimate to be
$M_{V,c}~$ = --4.8 to --5.4, respectively, for an $E(B-V)~$ in the
range 0.0 to 0.2. This star is therefore most likely a supergiant
F-star.

\begin{figure}
\begin{center}

{\epsfxsize 0.99\hsize
\epsffile{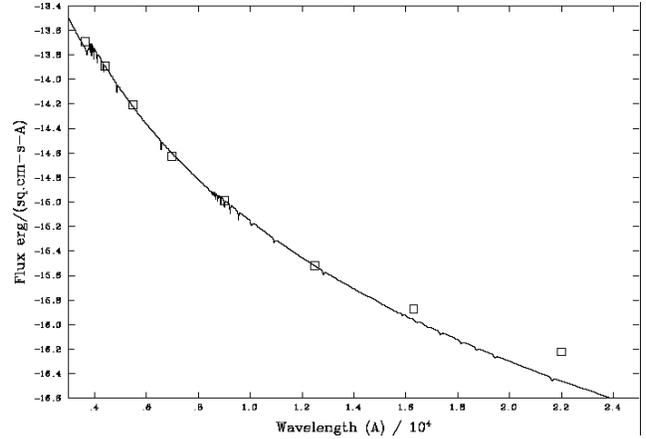}
}\end{center}
\caption{Photometric data points together with the best-fit Kurucz model for 
a B1-B2 III-V star ($\rm T_{eff} = 22000 \pm2000 K; log{\it g} = 4.0)$.}
\end{figure}

The degree of reddening seems consistent with the column density, $N_H$,
determined from the {\it ROSAT} observations. 
An $E(B-V)$ of 0.15 implies a column
of $N_H  = 1.4 \times 10^{21} \rm atoms~ cm^{-2}$, which compares favourably
with the value obtained from the X-ray spectrum (Section 2), namely 
$2.3 \pm1.1 \times 10^{21}\rm atoms~ cm^{-2}$. These column densities are all 
much larger than the column density to the SMC ($6 \times 10^{20} \rm 
cm^{-2}$), implying a local source of reddening consistent with a
circumstellar disk around the Be star.

In Section 2 we estimated the unabsorbed 0.1--2.0 keV X-ray flux as
\ap7.5 $\times 10^{-13}~\rm ergs~cm^{-2}~s^{-1}$. Using the same
power-law model, we calculate the unabsorbed 2-10 keV flux as \ap 3.2
$\times 10^{-13}~\rm ergs~cm^{-2}~s^{-1}$, which we use to determine
the luminosity ratio, $L_{X} (2-10keV)$ : $L_{opt} (3000-7000\AA)$, the
ratio used by Bradt \& McClintock (1983) in their survey of X-ray
source optical counterparts. Our estimate of $f_{opt, unabs}$ \ap 2
$\times 10^{-11}~\rm ergs~cm^{-2}~s^{-1}$ for Star A comes from the
optical spectrum (Fig. 4), where we integrated the flux over the
interval 4000-7000\AA, and extrapolated linearly to 3000\AA.  The
derived $L_{X}/L_{opt}$ (= $f_{X,unabs}/f_{opt, unabs}$) ratio of \ap
0.01--0.02 is very typical of HMXRBs, and the more active Be/X-ray
binaries. The ratio is some 4 orders of magnitude greater than that
expected from pure coronal emission from early type stars (Caillault
\& Helfand 1985), and consistent with accretion powered X-ray
emission. The inferred luminosity of \rx, $L_{X}$ (2-10keV) = 1.6$ \times
10^{35}$ ergs s$^{-1}$, is lower than X-ray ``high-state''
luminosities often seen in transient Be/X-ray systems, which are
typically in the range $\rm 10^{38}-10^{39}~ergs~s^{-1}$.  The value
is far more typical of the quiescent luminosity produced by stellar
wind accretion (Davidson \& Ostriker 1973).

It is interesting to note that \rx does not appear in the {\it ROSAT}
HRI survey of the SMC (Cowley et al. 1997). An upper limit of 0.001
cts s$^{-1}$ was placed on the flux, whereas it was expected to be
0.007 cts s$^{-1}$ based on the {\it Einstein} count rate reported in
Wang and Wu (1992). This clearly establishes the source as variable,
since it was observed by both {\it Einstein} and the {\it ROSAT} PSPC
(during the RASS). 

\rx is clearly a transient source, with excellent {\it bona fides} for a
Be/neutron star system. The X-ray, optical and IR variability (at
least for one candidate), combined with
the 92 s X-ray pulse period (Corbet et al. 1997),
convincingly demonstrates that \rx is a Be/neutron star binary system,
though we are unable to distinguish between the two possible optical
counterparts at this stage. Only if correlated optical/X-ray
variability is seen, or a much more accurate X-ray position determined,
will this point be resolved. 
This makes it the $5^{th}$ such system discovered in the SMC, following the 
discovery of RX J0117.6--7330 (Clark, Remillard and Woo 1997 and Coe
et al 1998).

We also note that a significant number of X-ray pulsars have recently 
been discovered in this region of the SMC. In addition to the source RX 
J0051.8-7231, from which 8.9-second pulsations were reported by Israel et 
al. (1997), the sources AX J0051-722 (91 second pulsations, Corbet et al. 
1998), 1WGA J0054.9-7226 (59 second pulsations, Marshall et al. 1998) 
and RX J0049.1-7250 (75 second pulsations, Yokogawa and Koyama 1998) have all 
been discovered in recently within a 40 arcminute region of the bar 
of the SMC. We have identified the optical counterparts to these X-ray 
pulsars to be SMC Be stars through similar optical programs as described 
here (Stevens, Coe and Buckley, 1999).

It is likely that the number of recent discoveries is simply due to 
the intensity of observations of this region following the discovery of 
the first pulsar. Future observations of the whole of the SMC with the 
new generation of X-ray telescopes should reveal whether this region 
really does contain an anomalous high density of X-ray pulsars.

\subsection*{Acknowledgments}
We are grateful to the very helpful Sutherland staff at the SAAO for their 
support during these observations. Dr Hartmut Winkler (Vista University, 
Soweto) kindly provided a spectrum of \rxj. The {\it Shara Focal Reducer} is 
on loan from Dr Michael Shara (STScI).

JBS is in receipt of a Southampton University studentship.

\end{document}